\documentclass[aps,prb,twocolumn,showpacs,floatfix,a4paper]{revtex4-2}
\usepackage{graphicx}
\usepackage[intlimits]{amsmath}
\usepackage{color}

\begin{document}

\title{Randomly Pruning the Sachdev-Ye-Kitaev model}
\author{Richard Berkovits}
\affiliation{Department of Physics, Jack and Pearl Resnick Institute, Bar-Ilan University, Ramat-Gan 52900, Israel}

\begin{abstract}

  The Sachdev-Ye-Kitaev model (SYK) is renowned for its short-time chaotic behavior, which plays a fundamental role
  in its application to various fields such as quantum gravity and holography. The Thouless energy,
  representing the energy scale at which the universal chaotic behavior in the energy spectrum ceases,
  can be determined from the spectrum itself. When simulating the SYK model on classical or quantum computers, it is advantageous
  to minimize the number of terms in the Hamiltonian by randomly
  pruning the couplings. In this paper, we demonstrate that even with
  a significant pruning, eliminating a large number of couplings, the chaotic behavior persists up to short time scales.
  This is true even when only
  a fraction of the original $O(L^4)$ couplings in the fully connected SYK model, specifically $O(KL)$, is retained.
  Here, $L$ represents the number of sites, and $K\sim 10$. The properties of the long-range energy scales,
  corresponding to short time scales, are verified through numerical singular value decomposition (SVD) and
  level number variance calculations.
  
\end{abstract}

\maketitle

\section{Introduction}

Recently we have witnessed the first use of a quantum computer
in order to calculate the dynamics of the Sachdev-Ye-Kitaev (SYK) model
\cite{jafferis22}. The SYK model has garnered significant attention as a model for holography and for
its ability to provide insights into the behavior of strongly interacting quantum systems \cite{kitaev15,sachdev15,maldacena16,jensen16}.
In this calculation \cite{jafferis22}, the primary objective was to implement a teleportation protocol that emulates
the dynamics of a traversable wormhole on the quantum computer. However, due to the current technical constraints
of quantum computers, the calculation was limited to a small number of qubits, and a pruned Hamiltonian was utilized,
discarding a majority of the coupling terms. To determine which terms to eliminate and to establish the amplitudes
of the remaining terms, a machine learning algorithm was employed. The goal was to preserve the chaotic behavior of the SYK model,
even in the sparse case. The success of this approach is currently a subject of debate and ongoing discussion \cite{kobrin23}.

The idea of simplifying the classical computation involved in the SYK model by removing certain coupling terms
was proposed in several previous studies \cite{xu20,garcia21,caceres22,tezuka23}. The approach involved pruning the coupling
terms either randomly or by ensuring that each site is coupled to a fixed fraction of other sites. In 
Ref. \cite{tezuka23}, the randomness of the couplings was further constrained to be either plus or minus a constant.
Remarkably, even with a significant reduction in the number of coupling terms, the chaotic dynamics of the SYK model persist.

In the SYK model, which consists of $N$ Majorana particles (or $N/2$ complex Fermions) and involves four Majorana (two-particle) interactions,
there are approximately $\binom {N}{4} \sim N^4$ coupling terms. Compared to the size of the Hilbert space, which is on the order
of $\binom {N/2}{N/4} \sim 2^{N/2}$, the number of coupling terms in this model is relatively small. However, as demonstrated in the
aforementioned references \cite{xu20,garcia21,caceres22,tezuka23}, even if a number of coupling terms on the order of $O(N)$ are retained,
the model continues to exhibit predominantly chaotic behavior.

  While much of the initial motivation for pruning the SYK model stemmed from simulating SYK on a quantum computer
  with the hope of gaining insights into properties of quantum gravity models, pruning the SYK also gives rise to broader
  questions related to deviations from chaotic behavior. It is well-known that realistic systems deviate from RMT
  at an energy scale known as the Thouless energy, $E_{Th}$, or at times shorter than the Thouless time,
  $t_{Th}=\hbar/E_{Th}$ \cite{mehta91,altshuler86}. The physical interpretation of the Thouless time for condensed
  matter metallic systems is straightforward, representing the time a particle needs to diffuse to the boundary
  of the system. This duration depends on the dimension and diffusion constant of the particle. For the standard SYK model,
  where each site is coupled to all the other sites, the energy spectrum follows RMT predictions up to the largest
  scales \cite{jia20,berkovits23}. 

  Therefore, if one aims to replace the standard SYK model with a pruned SYK model while preserving RMT behavior at very short
  time scales (large energies), a challenge arises to avoid introducing a spurious energy scale due to large sample-to-sample
  fluctuations. To mitigate the influence of these fluctuations, we employ a twofold approach. First, we transition to the
  unitary SYK model. Besides simplifying the implementation of SYK on both quantum and classical computers \cite{tezuka23},
  the constant amplitude of the couplings diminishes these fluctuations. Additionally, we employ an unfolding method, specifically
  Singular Value Decomposition (SVD), which has been demonstrated to enhance unfolding for the standard SYK model \cite{berkovits23}.

The most natural measure for
investigating the behavior of the energy spectra on these large energy scales is the
variance of the number of levels as function
of the size  of an energy window $E$, denoted by
$\langle \delta^2 n(E) \rangle$ (where $\langle \ldots \rangle$
represents an ensemble average and $n(E)$ is the number of levels within
$E$) \cite{mehta91,altshuler86}. A departure from the RMT behavior,
where  $\langle \delta^2 n(E) \rangle$ is proportional to $\ln(\langle n(E) \rangle)$,
above a certain
value of $E$ indicates that RMT holds only up to that energy scale.

As pruning becomes more severe, one expects the onset of chaotic behavior to occur at later times.
Consequently, deviations from the behavior predicted by Random Matrix Theory (RMT) should emerge at
lower energy scales. This culminates in a non-universal behavior at the single-level scale, implying a
complete loss of the RMT behavior in the case of extreme pruning.

Determining the energy scale at which the system deviates from RMT prediction, $E_{Th}$,
is not straightforward. As discussed it requires a careful unfolding process to
remove long-range average level spacing modulations, such as those arising from the band structure.
Similar challenges arise when attempting to identify the Thouless time through the ramp time in the
spectral form factor \cite{tezuka23}.

In this paper, we investigate the behavior of a specific SYK model as the degree of pruning varies.
In Section \ref{s1}, we define the complex SYK model and describe the procedure for random pruning.
Additionally, we introduce a new distribution of complex coupling terms in which the amplitude is constant,
but the phase is randomized. This distribution aims to minimize sample-to-sample fluctuations.
In Section \ref{s2}, we present various measures of chaotic behavior, including ratio statistics, level number variance,
and the Scree plot obtained through singular value decomposition (SVD). These measures are computed for
the unpruned SYK model, both for the standard unpruned and unitary unpruned SYK models,
serving as a baseline to establish the expected chaotic behavior.
Section \ref{s3} focuses on studying the impact of pruning on the density of states and the onset of
chaotic behavior. It is shown that for a not too drastic removal of couplings the pruned SYK system
retains its chaotic behavior at energy scales comparable to the fully connected model.
Finally, in Section \ref{s4},
we analyze the results obtained and discuss their implications for simulating the
SYK model using classical or quantum computers.

%}}}}}}}}}}}}}}}}}}}}}}}}}}}}}}}}}}}}}}}}}}

\section{Pruned complex SYK model}
\label{s1}

The complex spinless Fermions version of the
SYK model is given by the following Hamiltonian:
%modelR. Berkovits, Phys. Rev. B {\bf 107}, 035141 (2023)
\begin{eqnarray} \label{syk}
  \hat  H=
%  \sum_{i}^{L}\epsilon_i \hat c_i^\dag \hat c_i +
  \sum_{i>j>k>l}^{L} V_{i,j,k,l} \hat c_i^\dag \hat c_j^\dag \hat c_k \hat c_l,
%\\ \nonumber
%\hat  H= \frac{1}{2}\sum_{i,j} J_{i,j} \hat \chi_i^\dag \hat \chi_j +
%\sum_{i,j,k,l} \frac{1}{4!} J_{i,j,k,l}
%\hat \chi_i^\dag \hat \chi_j^\dag \hat \chi_k \hat \chi_l 
\end{eqnarray}
where  $\hat c_i^\dag$ is the creation operator of a Fermion on site $i$, and $L$ is the number of sites
and $L/2$ is the number of Fermions.
The couplings, $V_{i,j,k,l}$, are complex numbers where the real and imaginary components are traditionally
independently drawn from an identical Gaussian or box distribution, which will be denoted here as standard SYK.
We introduce an additional simplification by
drawing just a single random variable, $\theta_{i,j,k,l}$, from a uniform distribution between $0$ and $2\pi$,
denoted as the unitary SYK model.
This simplification yields $V_{i,j,k,l} = \exp(i \theta_{i,j,k,l})/(2L)^{3/2}$,
essentially a unitary-like distribution, which unlike
the Gaussian or box distribution has a constant amplitude. As we shall see, even
though these complex couplings are determined by a single random variable, they fully reproduce the chaotic
behavior observed in the SYK model. It is important to note that maintaining the complex
nature of $V_{i,j,k,l}$ is crucial. If one were to change it to a real random number
defined by a single random variable,
the model behavior would not exhibit the usual properties of the SYK model.

The pruning process involves randomly setting a fraction of the couplings, $V_{i,j,k,l}$, to zero.
As mentioned in the introduction, there are approximately $\binom {L/2}{4} \sim L^4$ couplings in total.
We randomly remove a majority of these couplings until only $KL$ coupling terms remain. To ensure the removal
process does not reduce the number of couplings on a site below $K-2$, we impose a constraint. This constraint
prevents the removal of a coupling if it would lead to a reduction in the number of couplings below $K$
on that particular site. As will be discuss further on, this constraint plays a role in smoothing the
density of states.

\section{Following RMT Behavior: Unpruned SYK Model}
\label{s2}

In order to set the stage for the influence of pruning of the couplings on the properties of SYK
models, it is important to present the properties of the unpruned SYK model. 
Throughout the paper we will consider a system of size $L=16$ corresponding to a matrix size
$\binom {L}{L/2} = 12870$ with $\binom {2L}{4}=35960$ couplings when no couplings are cut.
All statistics were performed on ensembles of $M=4000$ realizations of disorder.

\begin{figure}
\includegraphics[width=10cm,height=!]{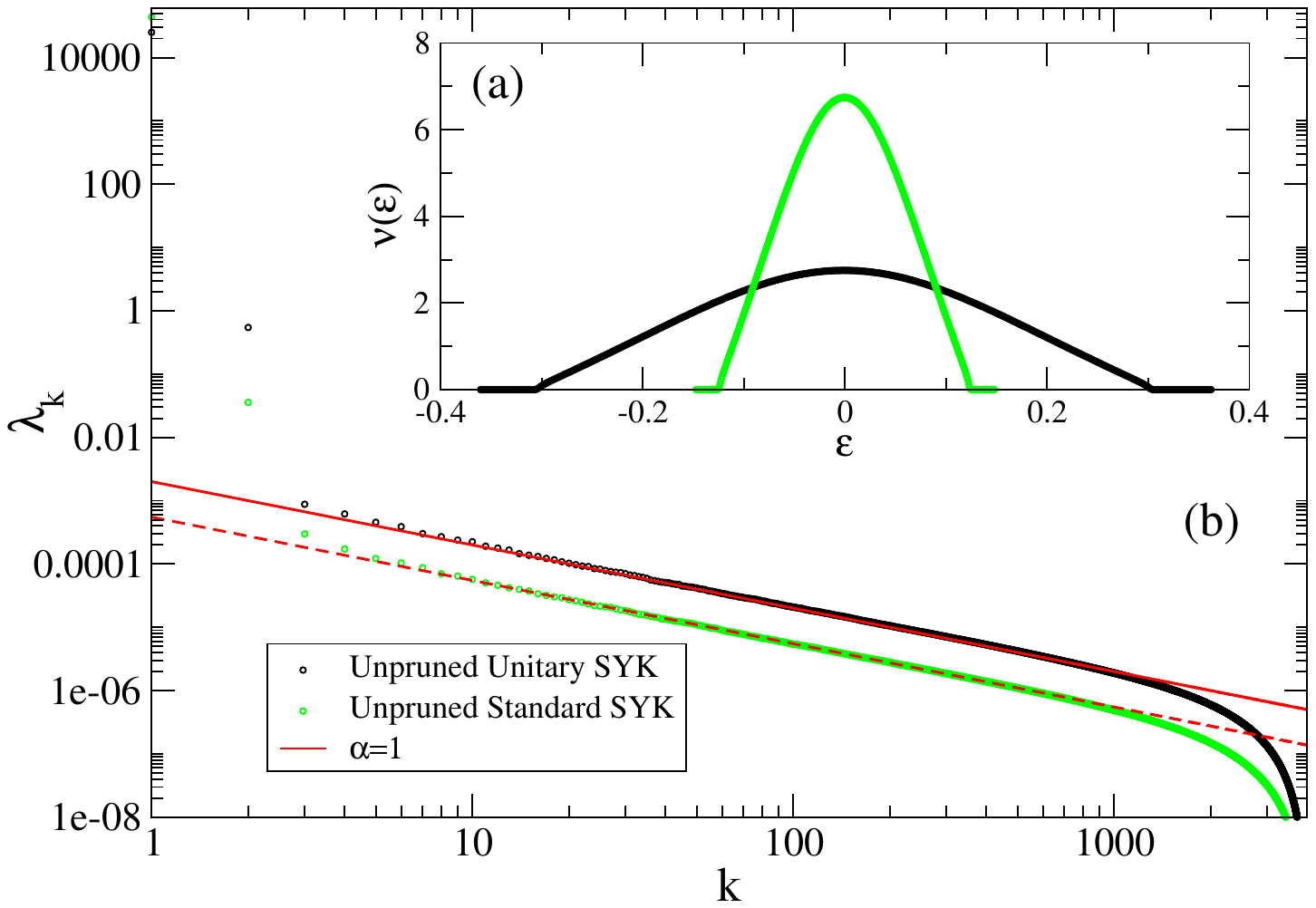}
\includegraphics[width=10cm,height=!]{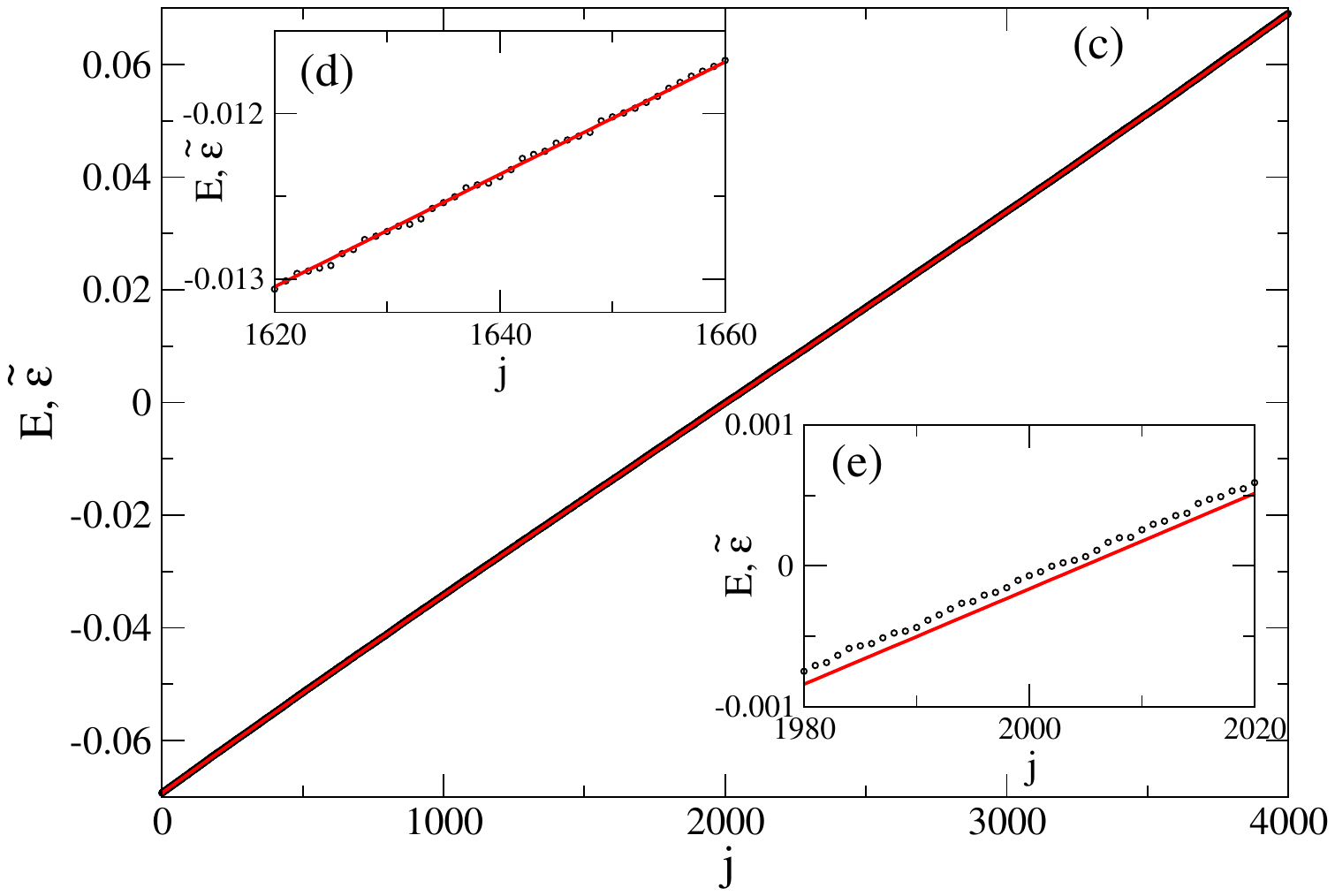}
\caption{\label{fig1}
  (a) The density of state $\nu(\varepsilon)$ as function of the energy
  for the unpruned unitary and standard SYK model.
    (b) The SVD amplitudes squared $\lambda_k$ vs. the mode number $k$. The unprunded unitary SYK results corresponds
  to the black symbols, while the unpruned standard SYK is represented by the green symbols. For both cases, the two lowest
  amplitudes corresponding to $k=1,2$ are orders of magnitude larger than the rest
  which follow $\lambda_k=k^{-\alpha}$.
    (c) The $4000$ levels straddling the middle of the band for the first realization of the unpruned unitary model.
    black circles correspond to the calculated eigenvalues $E_j$, while the red curve to
    $\tilde \varepsilon^1_i=\sum_{k=1}^2 \sigma_k U_{ik}V^T_{1k}$. (d,e) A zoom into smaller regions of the  eigenvalues.
}
\end{figure}

The first
feature we consider is the density of many-particle states $\nu(\varepsilon)$.
As can be seen in the inset of Fig. \ref{fig1} the density of states of the SYK model
with the unitary-like couplings is similar to the density of states
depicted in Ref. \onlinecite{garcia16} for the $N=32$ Majorana SYK model.

Next we consider the ratio statistics of the energy levels which
describe the short energy scale universal statistics. The
ratio statistics \cite{oganesyan07} are defined by:
\begin{eqnarray} \label{ratio}
r &=& \langle \min (r_n,r_n^{-1}) \rangle,
\\ \nonumber
r_n &=& \frac {E_n-E_{n-1}}{E_{n+1}-E_{n}},
\end{eqnarray}
where $E_n$ is the n-th eigenvalue of the Hamiltonian
and $\langle \ldots \rangle$ is
an average over an ensemble of different realizations of disorder and $4000$ of the eigenvalues
around the quarter point of the band. The reason for
avoiding the middle of the band would become clear in the next section.
For the Wigner GOE distribution expected for the $L=16$ \cite{garcia16,you17,li17}, $r_s \cong  0.5307$
\cite{atas13}, which is exactly the value obtained here.

For long range properties of the spectrum a useful measure
is the variance of the number of levels as function
of the size  of an energy window $E$, denoted by
$\langle \delta^2 n(E) \rangle$ (where $\langle \ldots \rangle$
represents an ensemble average and $n(E)$ is the number of levels within
$E$). As long as RMT behavior is followed the number variance
exhibits the Wigner GOE prediction $\langle \delta^2 n(E) \rangle =
(2/\pi^2) \ln(\langle n(E) \rangle) + 0.44$.
Departure from the RMT behavior is apparent above a certain
value of $E$, which quite naturally was identified with the Thouless energy \cite{altshuler86}.

It is important to emphasize that the functional form of the number variance for the Wigner GOE
was calculated assuming an average
level spacing of one. Therefore, before calculating the number variance, it is necessary to unfold
the spectrum so that the locally averaged level spacing is indeed equal to one. Due to its sparsity
and the limited number of independent random values of the coupling terms, which lead to strong
sample-to-sample fluctuations in the spectrum, the SYK model is extremely sensitive to the
unfolding procedure \cite{jia20,berkovits23}.

Consequently, employing the standard local unfolding procedure over an ensemble of different
realizations of disorder results in spurious deviations of the number variance from the RMT results.
Specifically, defining the averaged level spacing for the i-th level as
$\Delta_i = \langle E_{i+p}-E_{i-p}\rangle/2p$ (where $p$ is $O(1)$, and in this case chosen as $p=5$),
the unfolded spectrum for the j-th realization can be expressed
as $\varepsilon^j_i=\varepsilon^j_{i-1}+(E^j_i-E^j_{i-1})/\Delta_i$. This unfolding procedure
will significantly enhance the number variance.

Hence, a different unfolding procedure has been proven to correctly capture the sample-to-sample fluctuations
of the SYK model \cite{berkovits23}, and it will be used for the current study. This method is based
on the singular value decomposition (SVD) procedure
\cite{fossion13,torres17,torres18,berkovits20,berkovits21,berkovits22,rao22,rao23}. In this procedure,
the energy spectrum of the ensemble of different realizations is arranged as a matrix, where each row is
filled with $P$ consecutive eigenvalues of a given realization, while the rows correspond
to the $M$ different realizations.

In the SVD procedure, the $M \times P$ matrix $X$ is expanded as a sum over a series of amplitudes multiplied
by $M \times P$ matrices constructed by an outer product of two vectors of sizes $P$ and $M$. The vector
of size $P$ represents the averaged correction to the spectra over all realizations, while the $M$ terms
of the second vector provide a realization-dependent correction to the average. By plotting these amplitudes
from large to small (Scree plot), it is revealed that the largest amplitudes (usually just a few) are orders
of magnitude larger than the rest, while the following amplitudes obey a power law. These large amplitudes
depict the global behavior of the energy spectrum, while the power law corresponds to shorter-range properties.
This will be illustrated in the upcoming Fig. \ref{fig1}(c-e).

Explicitly, $X$ is decomposed to $X=U \Sigma V^T$, where
$U$ and $V$  are $M\times M$ and $P \times P$ matrices correspondingly,
while $\Sigma$ is a {\it diagonal} matrix of size $M \times P$ and rank
$r=\min(M,P)$. The $r$ diagonal elements of $\Sigma$
are the singular values amplitudes (SV) $\sigma_k$ of $X$.
The SV $\sigma_k>0$ 
and may be ordered by their magnitude
$\sigma_1 \geq \sigma_2 \geq \ldots \sigma_r$.
$X$ could be expressed as a series of matrices $X^{(k)}$,
where $X^{(k)}_{ij}=U_{ik}V^T_{jk}$  and $X_{ij}=\sum_k \sigma_k X^{(k)}_{ij}$.
The sum of the first $m$ modes gives an approximation
$\tilde X =\sum_{k=1}^m \sigma_k X^{(k)}$ to $X$ which represents
the minimal departure between the approximate energy
spectrum of all realizations, $\tilde X$, obtained using $m(M+P)$ independent
variables compared to the full energy spectrum which requires $MP$ variables. 
For the SVD unfolding the i-th level spacing of the j-th realization
$\Delta^j_i = (\tilde \varepsilon^j_{i+p}-\tilde \varepsilon^j_{i-p})/2p$, where
  $\tilde \varepsilon^j_i=\sum_{k=1}^l \sigma_k U_{ik}V^T_{jk}$, with $l$ corresponding to the first $l$ modes with amplitudes
which are orderes of magnitude larger than the folowing ones,
is realization dependent. 

In Fig. \ref{fig1}(a),
the density of states,  $\nu(\varepsilon)$, for the unpruned unitary (black symbols) and unpruned standard SYK (green) are presented.
The density of states for both cases are identical up to the averaged level spacing.
Fig. \ref{fig1}(b) displays the squared amplitudes, $\lambda_k=\sigma_k^2$,
ordered from large to small (known as a Scree plot)
for $L=16$ with $M=4000$
realizations and $P=4000$ eigenvalues around the quarter point of the many-particle band
for the unpruned unitary (black symbols) and unpruned standard SYK (green).
The amplitudes of the lowest modes ($k=1,2$) are orders of magnitude larger than those of the other modes,
thereby determining the non-universal behavior of the spectrum on a very large scale.
This is illustrated in Fig. \ref{fig1}c, where it is evident that the large-scale behavior of the energies
is accurately reproduced by modes $k=1,2$. However, it is important to note that these modes do not
capture local features, as revealed in Figs. \ref{fig1}d and \ref{fig1}e
For larger values of $k$, in both cases,
the majority of the modes exhibit a power-law behavior ($\lambda_k \sim k^{-\alpha}$)
with $\alpha=1$ for both cases as expected for Wigner-Dyson statistics \cite{fossion13,torres17,torres18}.
Referring to Ref. \onlinecite{berkovits23}, the energy at which the deviation from the universal
behavior becomes noticeable can be determined by examining the Scree plot. This deviation is observed
when the behavior deviates from the $k^{-1}$ power law, and the corresponding mode is denoted as $k_{Th}$.
Since $k_{Th}\sim 2$, the Thouless energy can be estimated as $E_{Th}=r \Delta/2 k_{Th} = 1000 \Delta$ \cite{berkovits22}.

\begin{figure}
\includegraphics[width=10cm,height=!]{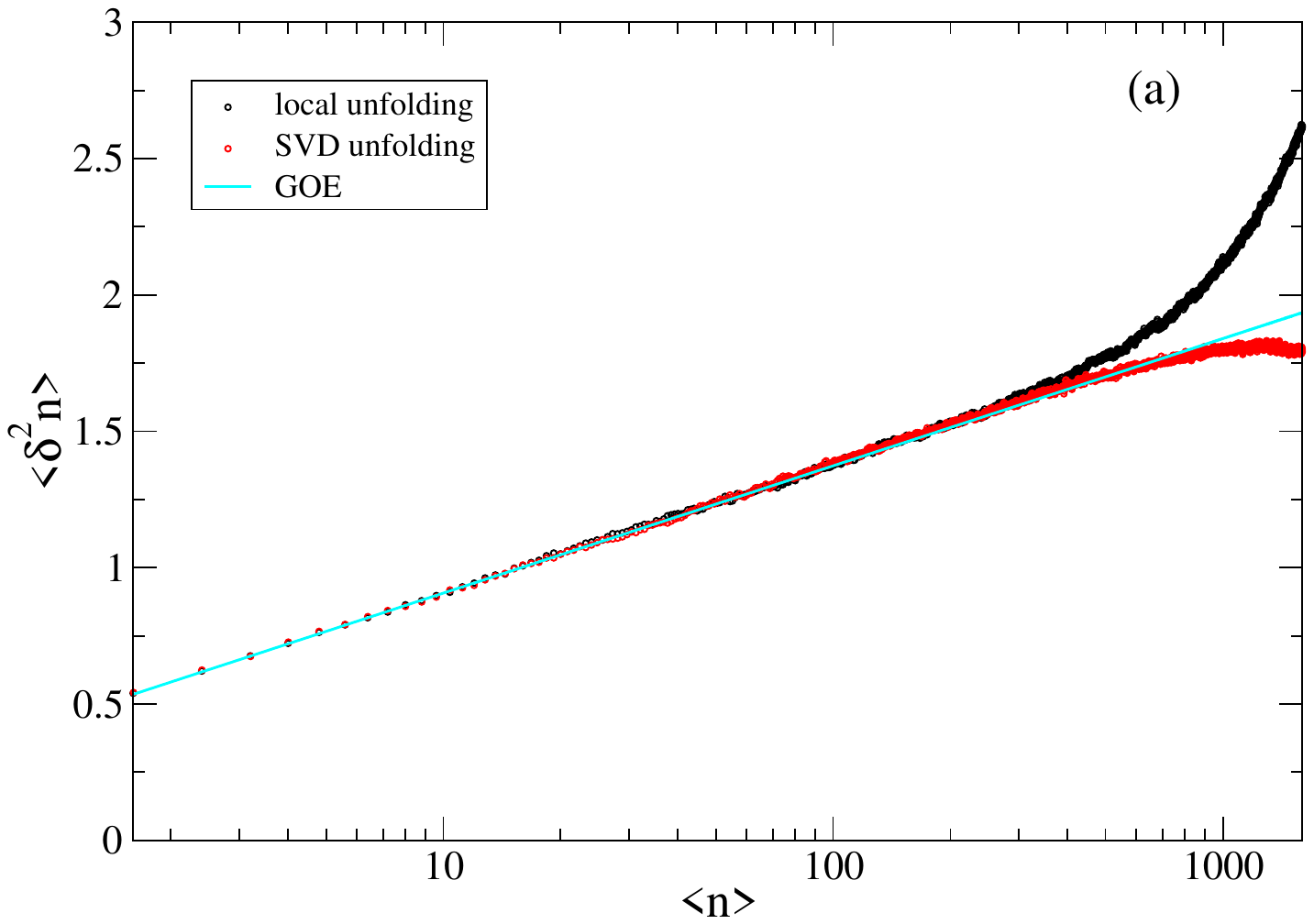}
\includegraphics[width=10cm,height=!]{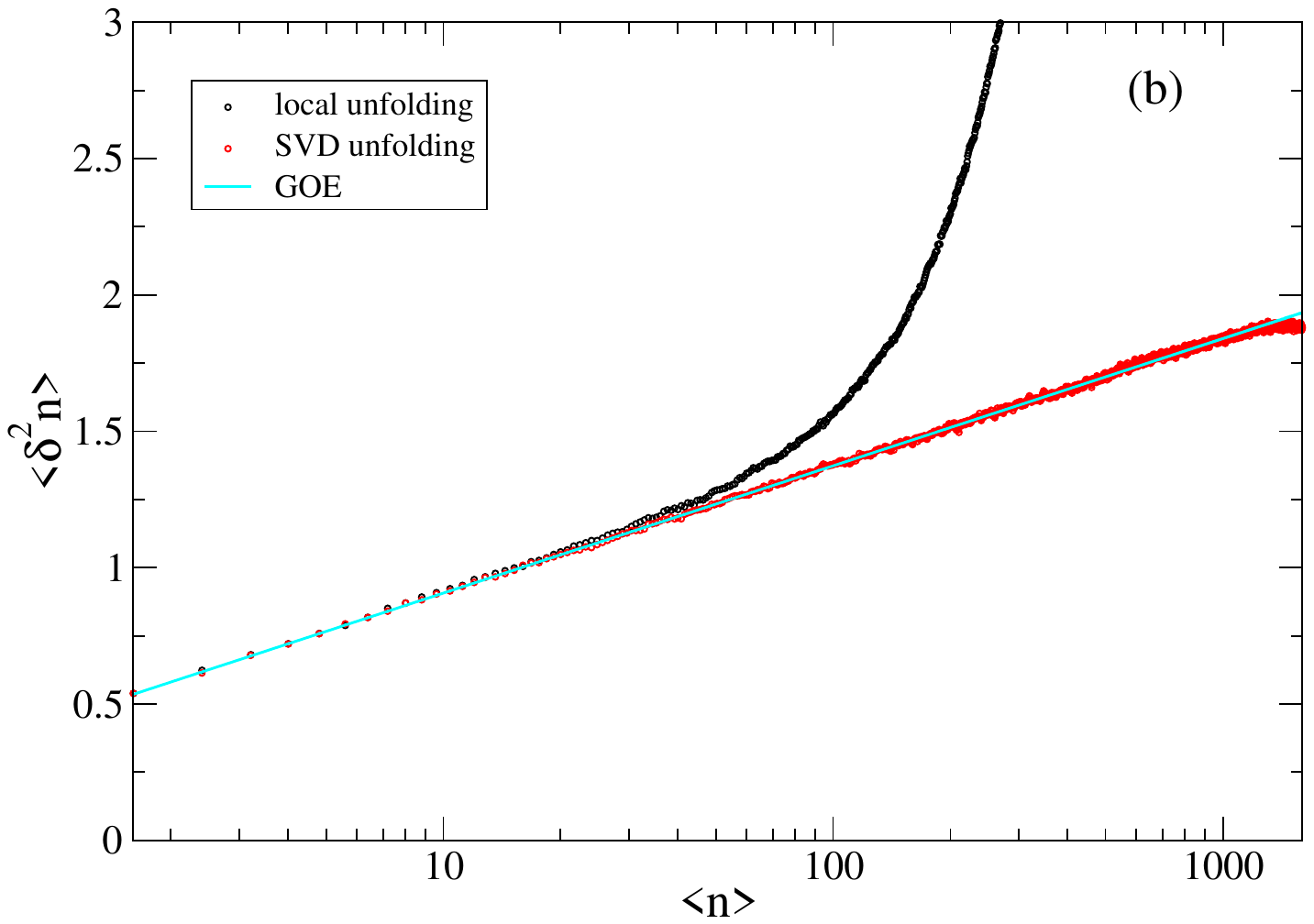}
\caption{\label{fig2}
 The level number variance, $\langle \delta^2 n(E) \rangle$, as
 function of the energy window $E$, (a) the unprumed unitary SYK model,
 (b) the unpruned standard model.
 The black symbols
  corresponds to the variance
  with local ensemble unfolding, while the red symbols to the SVD
  unfolding. The cyan 
  line is the GOE prediction $\langle \delta^2 n(E) \rangle =
  (2/\pi^2) \ln(\langle n(E) \rangle) + 0.44$.
}
\end{figure}

  Fig. \ref{fig2} illustrates the number variance using both unfolding methods. An intriguing observation emerges:
  unlike the standard unpruned SYK model, where the number variance strongly depends on the unfolding method
  (see Fig. \ref{fig2}(b)) \cite{berkovits23}, the difference is less pronounced for the unpruned unitary model
  (Fig. \ref{fig2}(a)). Nonetheless, employing the SVD unfolding noticeably enhances the fit to RMT predictions.
  It is noteworthy that RMT predictions align with expectations for any unpurned SYK model.
The Thouless energy
is estimated by considering the departure from the GOE predictions. Specifically, for the locally unfolding method,
the estimated Thouless energy is $E_{Th}=400 \Delta$, whereas for the Gaussian distribution SYK model it is
$E_{Th}=20 \Delta$ \cite{berkovits23}.
The SVD unfolding method deviates around $E_{Th}=800 \Delta$ (in reasonable agreement with a direct estimation using $k_{Th}$).
Beyond this point, the number variance tends to smaller values, a behavior consistent with observations from
other unfolding methods \cite{jia20}. The disparity between the standard SYK and unitary SYK models can be
attributed to the fact that the unitary random coupling model features couplings with random phases but constant amplitudes,
resulting in reduced sample-to-sample fluctuations.

  Therefore, the unpruned unitary SYK model exhibits chaotic behavior on large energy scales corresponding to short times,
  typically on the order of the bandwidth. As demonstrated in Ref. \cite{cotler17}, the spectral form factor of the standard
  SYK model replicates the RMT spectral form factor even for short times. Since the spectral form factor contains
  the same information as the energy spectrum, the same RMT behavior should be observed for $\langle \delta^2 n(E) \rangle$
  or the Scree plot at large energy scales. In the context of quantum gravity models, RMT measures of the spectrum serve as
  somewhat simpler indicators for more complex measures, such as the two-point correlation function of a Hermitian operator
  \cite{papadodimas15, cotler17, saraswat22}.
    Investigating whether distinct measures reveal varying facets of quantum chaos and their ability to capture the multifaceted
    nature of quantum chaos across different systems is a dynamic field of research. Researchers are actively engaged
    in unraveling the nuanced connections between diverse measures \cite{lewis19,huh21,brenes21,garcia-mata23}.
  Thus, accurately capturing the RMT behavior of the pruned SYK model across all
  energy scales becomes a crucial predictor for the validity of replacing the standard SYK model with a pruned one.

\section{Pruned Unitary SYK's Departure from RMT Behavior}
\label{s3}

Given that reducing the number of couplings is crucial for achieving a realistic simulation
of the SYK model on a quantum computer, it is essential to estimate the extent to which
couplings can be dropped while still preserving the unpruned SYK chaotic behavior on the
shortest time scales.

\begin{figure}
\includegraphics[width=10cm,height=!]{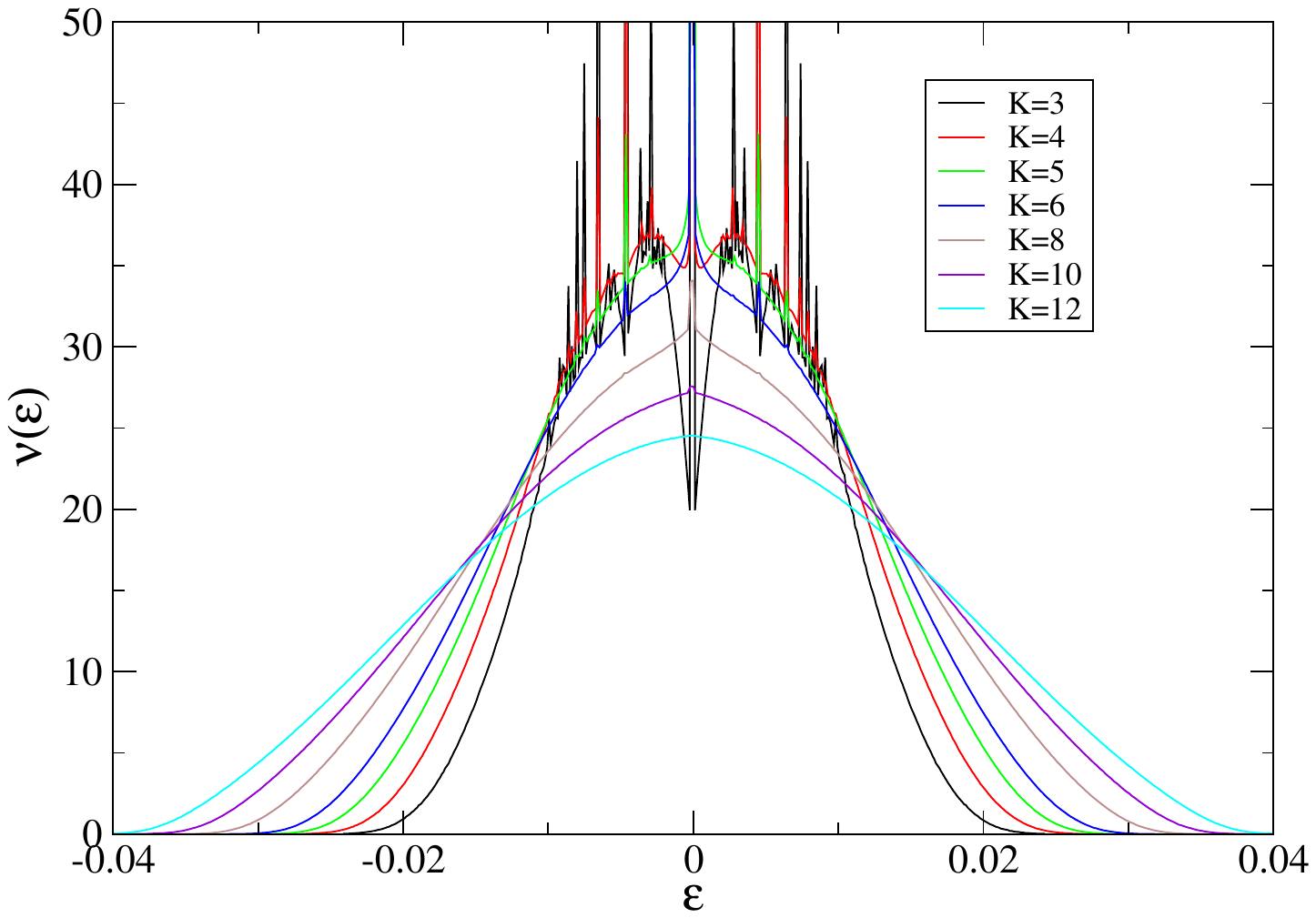}
\caption{\label{fig3}
  The density of states $\nu(\varepsilon)$ as a function of energy $\varepsilon$ is examined for the unitary
  SYK model with different degrees of pruning represented by $K$. For $K=12$, the density of states
  appears continuous and exhibits similarity (up to scaling) to the unpruned SYK model.
  However, as the pruning becomes more severe (lower values of $K$), $\nu(\varepsilon)$ develops additional
  structure and discontinuities, particularly near the middle of the energy band ($\varepsilon=0$).
  }
\end{figure}

To begin, we investigate the density of states as a function of energy (refer to Fig. \ref{fig3})
to gain insights into the impact of pruning on $\nu(\varepsilon)$. As the pruning becomes more severe,
indicated by a decrease in the value of $K$, the density of states no longer resembles the continuous
and well-behaved density observed in the unpruned SYK model (Fig. \ref{fig1}). However, as long as
the pruning is not too drastic ($K \geq 12$), $\nu(\varepsilon)$ remains continuous and lacks sharp features.
It retains a resemblance, albeit scaled, to the density of states observed in the upruned unitary SYK model.
On the other hand, for stronger pruning (with $K<12$), $\nu(\varepsilon)$ exhibits a greater number of
discontinuities.

These spikes in the density of states initially emerge near the midpoint of the energy band, specifically at $\varepsilon=0$.
For $K$ values down to $6$, a single spike at $\varepsilon=0$ represents the primary influence of pruning on
the density of states. This is why we focus on analyzing the long-range properties of the spectra away from the midpoint.
However, as the value of $K$ decreases further, these spikes begin to appear in other regions of the spectrum.
The origin of these spikes can be understood by considering the behavior observed in the density of states of
the bond percolation model \cite{berkovits96}. As the number of couplings decreases, more disconnected small clusters of sites emerge.
Zero energy peaks in the density of states have been observed in quantum network models exhibiting similar
connectivity \cite{yadav15,marrec17,bueno20}. Additionally, larger clusters are expected to manifest as peaks at distinct energy values.

\begin{figure}
\includegraphics[width=10cm,height=!]{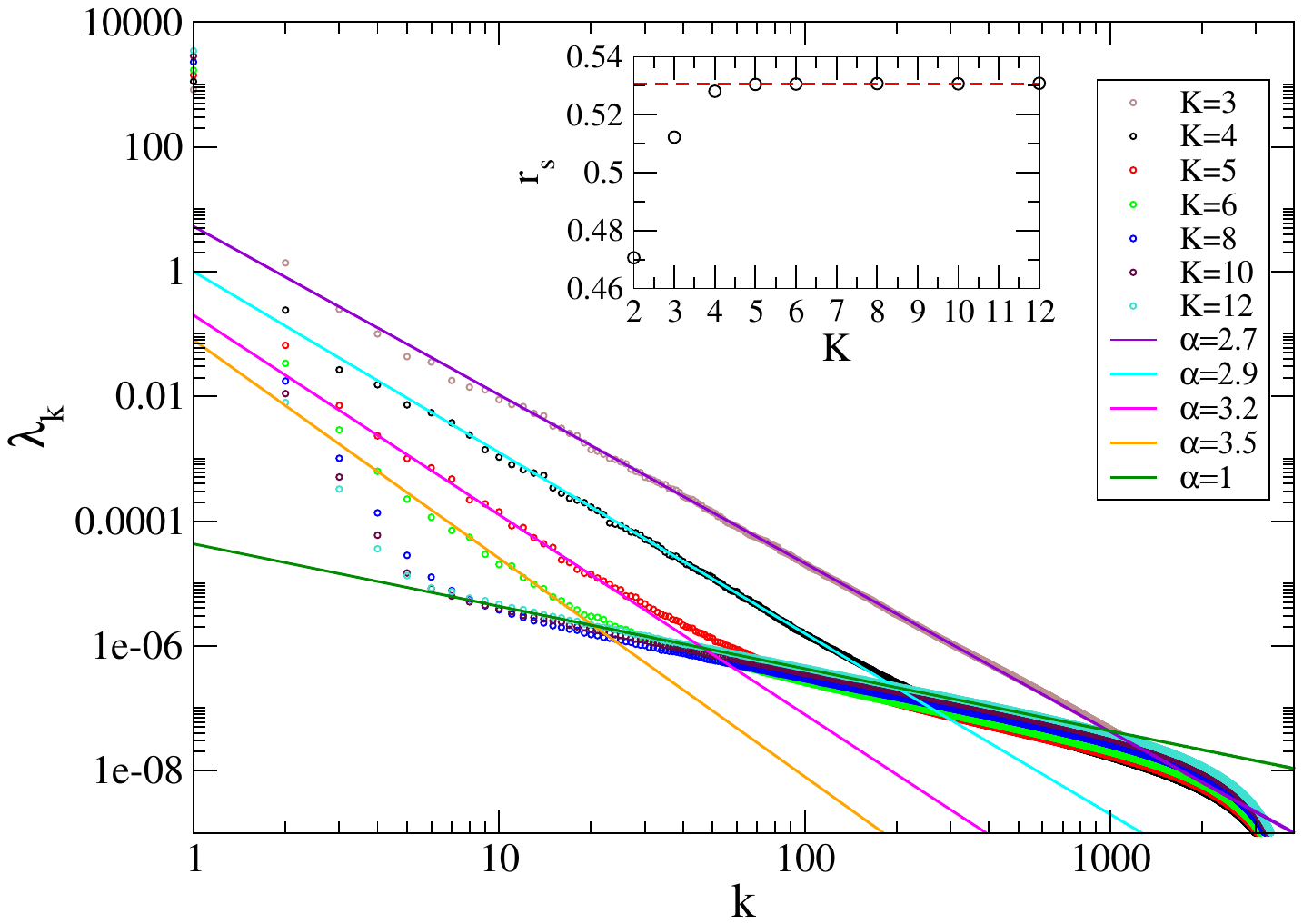}
\caption{\label{fig4}
  The plot showcases the squared SVD amplitudes, $\lambda_k$, plotted against the mode number $k$ (Scree plot),
  varying with the pruning parameter $K$ for the unitary SYK model. The lines in the plot represent power
  laws of the form $\lambda_k=k^{-\alpha}$.
  The predicted power law exponent for the GOE is $\alpha=1$. The inset displays the ratio statistics,
  $r_s$, also as a function of the pruning parameter $K$. The red dashed line in the inset corresponds to the
  results predicted by the GOE.
}
\end{figure}

The short range energy spectra properties , as measured by the ratio statistics are presented in
the inset of Fig. \ref{fig4}. It can be seen that down to $K=5$ the short range properties of
the pruned unitary SYK model follows RMT predictions at long times (short energy scales).
For extreme pruning ($K \leq 4$) the results deviate even for nearest neighbor statistics.

The short-range properties of the energy spectra, as assessed by the ratio statistics, are shown
in the inset of Fig. \ref{fig4}. It is evident that, until $K=5$, the pruned unitary SYK model exhibits
short-range properties consistent with the predictions of RMT at longer times (corresponding to shorter energy scales).
However, for extreme pruning ($K \leq 4$), the results deviate from RMT predictions, even in the case of
nearest-neighbor statistics.

The picture undergoes a dramatic transformation when we scrutinize the energy properties on a large scale.
Similar to our analysis of the classic unitary SYK model, we employ SV amplitudes to investigate the energy
scale at which pruned SYK models deviate from RMT. As expected, in cases of extreme pruning where even
the shortest scales do not adhere to RMT predictions, the answer becomes evident at any scale. In fact,
as depicted in the Scree plot illustrated in Fig. \ref{fig4}, for $K=3$, $\lambda_k \sim k^{-2.7}$ up to
the highest values of $k$ (i.e., corresponding to small energies). As the value of $K$ increases, a finite
region of modes emerges where $\lambda_k$ follows RMT. The mode at which SV amplitudes diverge from RMT
and transition to a different power-law dependence $\lambda_k \sim k^{-\alpha}$ with $\alpha>2$ can be
identified as $k_{Th}$. By examining Fig. \ref{fig4}, we observe that this pattern holds true for
$4 \leq K \leq 6$, with $k_{Th}(K=4)\sim 250$, $k_{Th}(K=5)\sim 100$, and $k_{Th}(K=6)\sim 50$, corresponding to
$E_{Th}(K=4)\sim 8 \Delta$, $E_{Th}(K=5)\sim 20 \Delta$, and $E_{Th}(K=6)\sim 40 \Delta$, respectively.
In the case of more moderate pruning, the number of modes is insufficient to identify a power-law behavior
for smaller modes. However, we can still determine the mode at which the amplitude deviates from the
GOE prediction, yielding $k_{Th}(K=8)\sim 10$, $k_{Th}(K=10)\sim 5$, and $k_{Th}(K=12)\sim 5$, corresponding
to $E_{Th}(K=8)\sim 200 \Delta$, $E_{Th}(K=10)\sim 400 \Delta$, and $E_{Th}(K=12)\sim 400 \Delta$,
respectively. Thus, as pruning becomes less extreme, the deviation from RMT is pushed to higher energies
(shorter times).

\begin{figure}
\includegraphics[width=10cm,height=!]{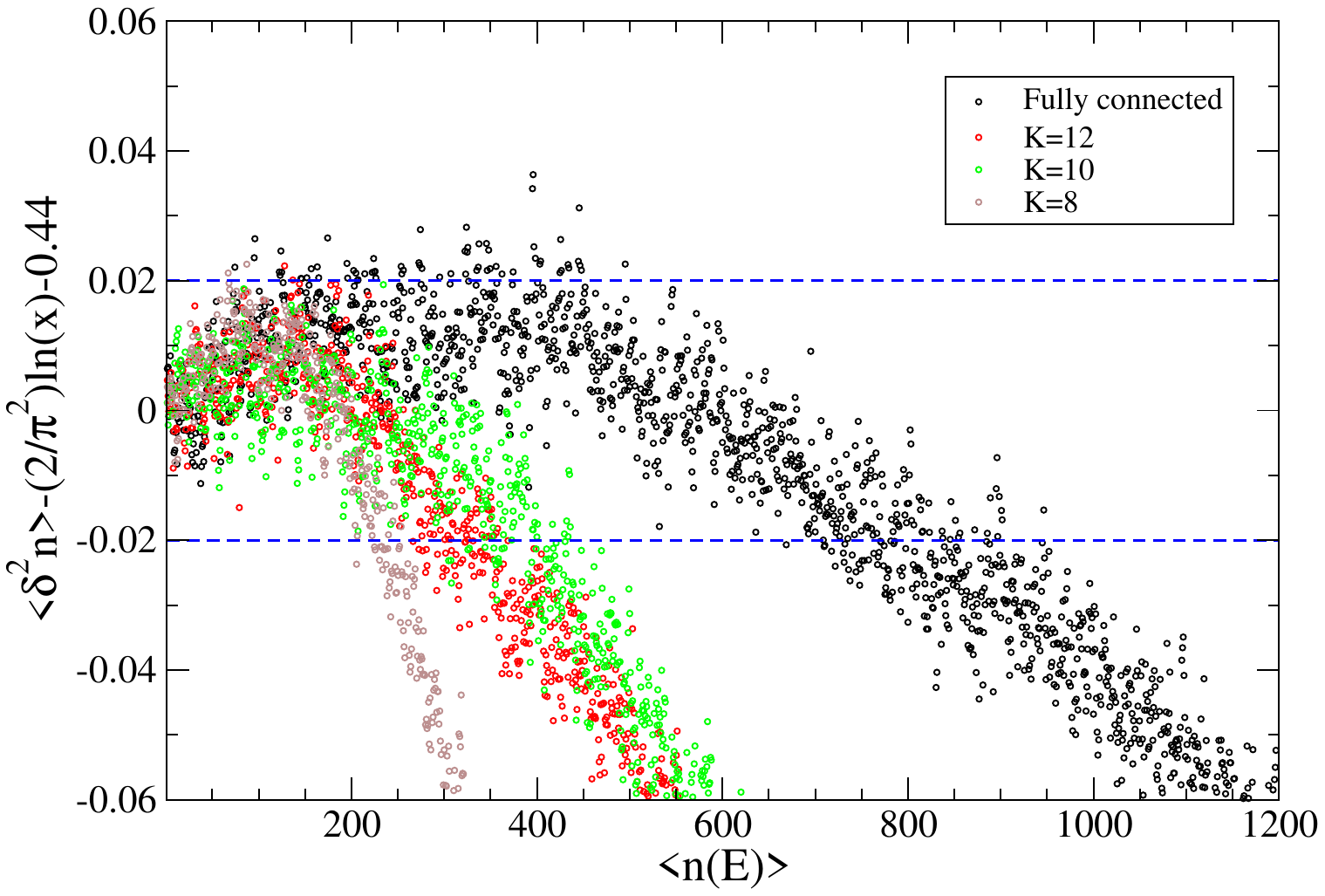}
\caption{\label{fig5}
  The plot illustrates the level number variance, $\langle \delta^2 n(E) \rangle$, as a function of the energy window $E$,
  utilizing the SVD unfolding technique. The symbols represent the numerical variances for different cases:
  the unpruned unitary SYK (black), and different pruned unitary models $K=12$ (red), $K=10$ (green), and $K=8$ (brown).
  By plotting
  $ \langle \delta^2 n(E) \rangle - (2/\pi^2) \ln(\langle n(E) \rangle) - 0.44$ (i.e., the variance with the GOE value subtracted),
  we observe that the variance, post-SVD unfolding, remains within the expected range of the GOE (indicated up to $\pm 0.02$
  by blue dashed lines) for $n(E) < n(E_{Th})$. This allows us to estimate the threshold energies as
  $E_{Th}=800 \Delta,400 \Delta,400 \Delta,250 \Delta$
  for the unpruned unitary case, and pruned unitary $K=12,10,8$ respectively.
}
\end{figure}

This hypothesis can be directly tested by examining the spectrum's level number variance after applying SVD unfolding.
The results are presented in Fig. \ref{fig5}, where the variance is plotted after subtracting the expected GOE behavior:
$\langle \delta^2 n(E) \rangle - (2/\pi^2) \ln(\langle n(E) \rangle) - 0.44$. For small energy windows,
the variance follows the GOE within a margin of $\pm 0.02$. However, as the energy increases, the correspondence
between the numerical results and the GOE ends. The deviation point from the GOE decreases as the pruning becomes more
aggressive. Specifically, the Thouless energy, $E_{Th}=800 \Delta,400 \Delta,400 \Delta,250 \Delta$ for the unpruned case and $K=12,10,8$
respectively, shows reasonable agreement with the Thouless energy obtained from $k_{Th}$. Therefore, the long-range behavior
of the unpruned SYK model is preserved when the couplings are pruned moderately (around $K \sim 10$).

\section{Discussion}
\label{s4}

Therefore, this study provides further confirmation of the key finding in previous investigations \cite{xu20,garcia21,caceres22,tezuka23},
namely, that the chaotic nature of the SYK model persists even under substantial pruning, where only $KL$ coupling terms are retained
instead of the original $O(L^4)$ terms. In contrast to the methods employed in the aforementioned studies, which utilized spectral
form factors, our approach involved analyzing the SVD amplitudes and utilizing the SVD unfolded level number variance.
Through these techniques, we determined the energy scale at which the energy spectrum deviates from the behavior expected by GOE,
thereby establishing the Thouless energy of the pruned SYK model.

While the overall broad picture emerging from this study aligns with the findings of previous investigations, there are some nuances
worth noting. One must distinguish between the previous studies focusing on Majorana particles and the current study, which investigates
a complex Fermion model when comparing the results of the two models. It is important to recognize that $N$ Majoranas correspond to $2L$ sites.
Consequently, when examining the results of retaining $KL$ couplings in the current study, they should be compared to $K_{M}N$ for the Majorana SYK models,
resulting in $K=2K_{M}$.

Notably, when comparing the estimates of the pruning limit for which chaotic behavior is preserved, the agreement between
the spectral form factor results \cite{xu20,garcia21,caceres22,tezuka23} for the Majorana models with $K_{M} \geq 4$ and our SVD findings of $K\geq 10$
is remarkable. However, there is a noticeable disparity when examining the pruning estimates presented in Ref. \onlinecite{tezuka23}.
In that study, $K_{M} \geq 4$ is applicable only to the dichotomic ($\pm 1$) distribution of couplings, while a higher value of $K_{M}$ is observed for a
Gaussian distribution of couplings, as used in \cite{xu20,garcia21,caceres22}. Similar to the discussion regarding
the unpruned SYK model in Sec. \ref{s2}, this disparity may be attributed to the stronger sample-to-sample
fluctuations in the Gaussian distribution, leading to a greater sensitivity of the apparent onset of chaotic behavior to the details of the unfolding procedure.

In the case of our model, which investigates the complex Fermion SYK, implementing the dichotomic distribution is not feasible since the couplings
must be complex to preserve the SYK symmetries. Therefore, we utilized a unitary-like distribution with constant coupling amplitudes. The obtained
value of $K$ is consistent with the previous results, further supporting the notion of preserving chaotic behavior as long as the pruning is not too radical.

In conclusion, our findings demonstrate that a pruned SYK model exhibits chaotic behavior at very short time scales,
even when only a relatively small number of couplings (on the order of the number of Majorana particles or sites) are retained.
However, it is important to note that once the number of remaining couplings becomes too low (less than $10L$ for the complex SYK model),
the behavior of the energy spectrum at large energy scales deviates from the expected chaotic pattern. In the case of random pruning,
the density of states displays additional peaks associated with disconnected clusters, which must be carefully handled.
By replacing the Gaussian distribution of couplings with a unitary-like distribution, we can reduce the sample-to-sample
fluctuations that affect the former approach, although it does not completely eliminate the need for careful unfolding.
These results highlight the significance of these considerations when simulating the SYK model using classical or quantum computers.

\end{document}